\documentclass[onecolumn,nofootinbib,showpacs,preprintnumbers,amsmath,amssymb]{revtex4-1}

\usepackage{amsfonts}
\usepackage{amsmath}
\usepackage{amssymb}
\usepackage{amstext}
\usepackage{amsthm}
\usepackage{graphicx}
\usepackage{dcolumn}
\usepackage{bm}
\usepackage{color}
\usepackage{float}

\begin{document}

\title{The Double Role of GDP in Shaping the Structure of the International Trade Network}

\author{Assaf Almog}
\affiliation{Instituut-Lorentz for Theoretical Physics,Leiden Institute of Physics, University of Leiden, Niels Bohrweg 2, 2333 CA Leiden (The Netherlands)}

\author{Tiziano Squartini}
\affiliation{IMT Institute for Advanced Studies, P.zza S. Ponziano 6, 55100 Lucca (Italy)}

\author{Diego Garlaschelli}
\affiliation{Instituut-Lorentz for Theoretical Physics, Leiden Institute of Physics, University of Leiden, Niels Bohrweg 2, 2333 CA Leiden (The Netherlands)}

\date{\today}

\begin{abstract}
The International Trade Network (ITN) is the network formed by trade relationships between world countries. 
The complex structure of the ITN impacts important economic processes such as globalization, competitiveness, and the propagation of instabilities.
Modeling the structure of the ITN in terms of simple macroeconomic quantities is therefore of paramount importance. 
While traditional macroeconomics has mainly used the Gravity Model to characterize the magnitude of trade volumes, modern network theory has predominantly focused on modeling the topology of the ITN. Combining these two complementary approaches is still an open problem.
Here we review these approaches and emphasize the double role played by GDP in empirically determining both the existence and the volume of trade linkages. Moreover, we discuss a unified model that exploits these patterns and uses only the GDP as the relevant macroeconomic factor for reproducing both the topology and the link weights of the ITN.

\end{abstract}

\maketitle

 \section{Introduction}

The bilateral trade relationships existing between world countries form a complex network known as the International Trade Network (ITN).
The observed complex structure of the network is at the same time the outcome and the determinant of a variety of underlying economic processes, including economic growth, integration and globalization. Moreover, recent events such as the financia l crisis clearly pointed out that the interdependencies between financial markets can lead to cascading effects which, in turn, can severely affect the real economy. International trade plays a major role among the possible channels of interaction among countries \cite{Kali1,Kali2,integration,Saracco}, thereby possibly further propagating these cascading effects worldwide and adding one more layer of contagion. Characterizing the networked worldwide economy is therefore an important open problem and modelling the ITN is a crucial step in this challenge, and has been studied extensively \cite{Serrano,Vespignani,Fagiolo1,Barigozzi,DeBene1,Pietronero,Sinha}.\\

Historically, macroeconomic models have mainly focused on modelling the trade volumes between countries. The Gravity Model, which was introduced in the early 60's by Jan Tinbergen \cite{Tinbergen2}, serves as a powerful empirical model that aims at inferring the volume of trade between any two (trading) countries from the knowledge of their Gross Domestic Product (GDP) and mutual geographic distance. Over the years, the model has been upgraded to include other possible factors of macroeconomic relevance, like common language and trade agreements, nevertheless GDP and distance remain the two factors with biggest explanatory power.
The gravity model can reproduce the observed trade volume between trading countries satisfactorily. However, at least in its simplest and most popular implementation, the model does not generate zero volumes and therefore predicts a fully connected trade network.
This outcome is totally inconsistent with the heterogeneous observed topology of the ITN, which serves as the backbone on which trades are made. More sophisticated implementations of the gravity model that do a llow for zero trade flows succeed only in reproducing the number of missing links, but not their position in the trade network, thereby producing sparser but still non-realistic topologies \cite{Fagiolo2,Fagiolo3}.\\

In conjunction with the traditional macroeconomic approach, in recent years the modelling of the ITN has also been approached using tools from network theory \cite{Diego1,Garlaschelli,Caldarelli,Fronczak,Bhattacharya}, among which maximum-entropy techniques \cite{Squartini2,Squartini3,Squartini4} have been particularly successful.
Maximum-entropy models aim at reproducing higher-order structural properties of a real-world network from low-order, generally local information, which is taken as a fixed constraint \cite{Wells,Bargigli,Musmeci,Caldarelli2}. Important examples of local properties that can be chosen as constraints are the \emph{degree}, i.e. the number of links, of a node (for the ITN, this is the number of trade partners of a country) and the \emph{strength}, i.e. the total weight of the links, of a node (for the ITN, this is the total trade volume of a country). Examples of higher-order properties that the method aims at reproducing are \emph{clustering}, which refers to the fraction of realised triangles around a node, and \emph{assortativity}, which is a measure of the correlation between the degree of a node and the average degree of its neighbours. 

These studies have focused on both binary and weighted representations of the ITN, i.e. the two representations defined by the \emph{existence} and by the \emph{magnitude} of trade exchanges among countries, respec tively.
In principle, depending on which local properties are chosen as constraints, maximum-entropy models can either fail or succeed in replicating the higher-order properties of the ITN. 
As an example, it has been shown that inferring a network topology only from purely weighted properties such as the strength of all nodes (i.e. the trade volumes of all countries) results in a trivial, uniform structure (almost fully connected and, thus, unrealistic) \cite{Random2}. This limitation is similar to the one discussed above for the gravity models, which aim at reproducing the pair-specific traded volumes exclusively, while completely ignoring the underlying network topology. By contrast, the knowledge of purely topological properties such as the degrees of all nodes (i.e. the number of trade partners of all countries), which are usually neglected in traditional macroeconomic models, turns out to be essential for reproducing the heterogeneous topology observed in the ITN \cite{Random1}. A combination of weighed and topological local properties allows to reconstruct the higher-order properties of the ITN with extremely high accuracy \cite{Squartini6}.\\

Despite the ability of the appropriate maximum-entropy models to provide a better agreement with the data with respect to gravity models, they do not in principle provide any hint on the underlying (macro)economic factors shaping the structure of the network under consideration. These models, in fact, assign ``hidden variables'' or ``fitness parameters'' to each country. These quantities arise as Lagrange multipliers involved in the constrained maximisation of the entropy and control the probability that a link is established and/or has a given weight. These parameters have, a priori, no economic interpretation. However, here we show that one can indeed find a macroeconomic iden tificatio n for the underlying variables defining the maximum-entropy models. This interpretation is supported by previous studies showing that both topological and weighted properties of the ITN are strongly connected with purely macroeconomic quantities, in particular the GDP.\\

In this paper we first focus on various empirical relations existing between the GDP and a range of country-specific properties. These properties convey basic but important local information from a network perspective. We also show that these relations are robust and very stable throughout different decades. We then illustrate how the GDP affects differently the binary and weighted representations of the ITN, revealing alternative aspects of the structure of this network. These results suggest a justification for the use of GDP as an empirical fitness to be used in maximum-entropy models, thus providing a macroeconomic interpretation for the abstract mathematical parameters defining the model t hemselves . Reversing the perspective, this result enables us to introduce a novel GDP-driven model \cite{Almog} that successfully reproduces the binary \emph{and} the weighted properties of the ITN simultaneously. The mathematical structure of the model explains the aforementioned puzzling asymmetry in the informativeness of binary and weighted constraints (degree and strength) \cite{Almog}. These results represent a promising step forward in the formulation of a unified model for modelling the structure of the ITN.

\section{Data} \label{data}

In this study we have used data from the Gleditsch database which spans the years 1950-2000 \cite{Gleditsch}, focusing only on the first year of each decade, i.e. six years in total. The data sets are available in the form of weighted matrices of bilateral trade flows $w_{ij}$, the associated adjacency matrices $a_{ij}$ and vectors of GDPs. There are approximately 200 countries in the data set covering the considered 51 years; the GDP is measured in U.S. dollars. 
\\

We have analysed this data set precisely because it has been the subject of many studies so far, focusing both on the binary and on the weighted representation of the ITN. This will allow us to compare the performance of our GDP-driven (two-steps) method with other reconstruction algorithms already present in the literature \cite{Squartini5}.

Trade exchanges between countries play a crucial role in many macroeconomic phenomena. As a consequence, it is fundamental to be able to characterize the observed structure of the ITN and its properties. More specifically, the ITN can be represented in two different ways, depending on the kind of information used to analyse the system: the first one concerns only the existence of trade relations and gives origin to the ITN \emph{binary representation}; the second one also takes into account the volume of the trade exchanges and gives origin to the ITN \emph {weighted representation}. While the binary representation describes the skeleton of the ITN, relating exclusively to the presence of trade relations, the weighted representation also accounts for the volume of trade occurring ``over'' the links, i.e. the weight of the link once it is formed. The two representations convey very important information regarding the ``trade patterns'' of each country and, most importantly, correspond to different trade mechanisms.\\
\begin{figure*}[]
\caption{Empirical cumulative distributions $P_>(\tilde{g})$ of the GDP rescaled to the mean, for different years. The curve is log-normal distribution fitted to the data.}
\centerline{\includegraphics[width=0.9\textwidth]{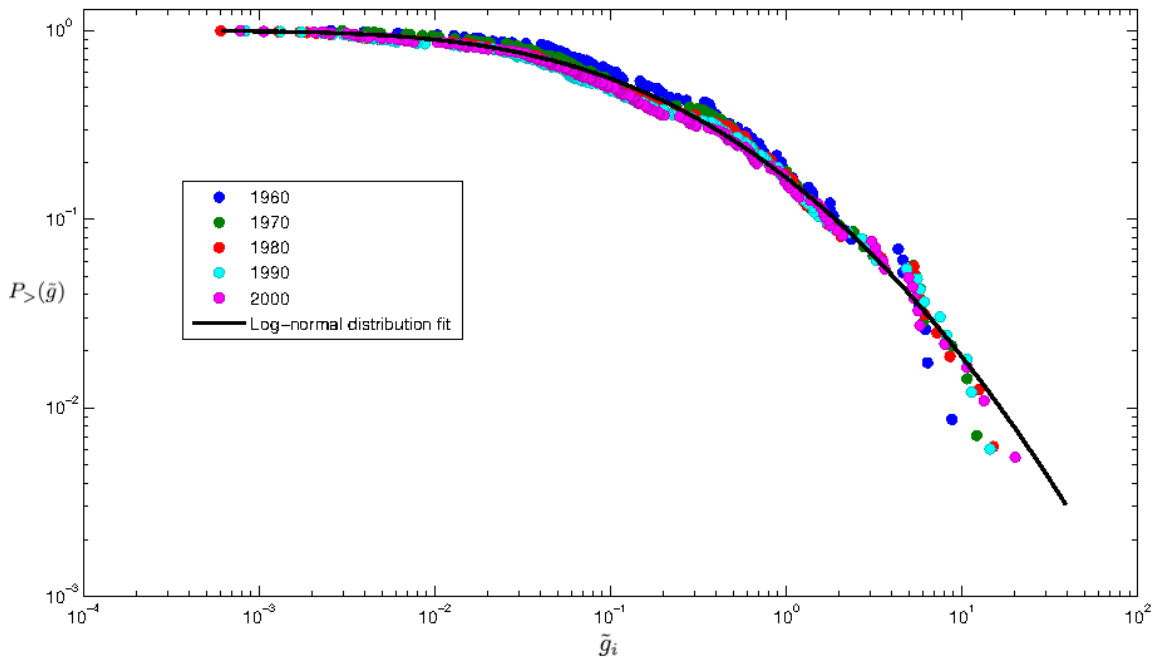}}
\label{fig1}
\end{figure*}

Traditionally, macroeconomic models have mainly focused on the weighted representation, because economic theory perceives the latter as being genuinely more informative than the purely binary representation: such models make use of countries gross domestic product (GDP), their geographic distance and any other possible quantity of (supposed) macroeconomic relevance to infer trading volumes between countries. The GDP is the most popular measure in the economic literature. Although it is generally used as a proxy to infer the evolution of many macroeconomic properties describing the weighted representation of the ITN (as the countries trade exchanges), here we will show that the GDP plays a key role not only to explain the ITN weighted structure, but also the emergence of its binary structure.\\ 

Let us start with an empirical analysis of the GDP. We first define new rescaled quantities of the GDP: $g_i$ and $\tilde{g}_i$
\begin{equation}
g_i\equiv \frac{\mbox{GDP}_i}{\sum_j\mbox{GDP}_j},\:\forall\:i~~~~~~~
\tilde{g}_{i}\equiv \frac{\mbox{GDP}_i}{\mbox{GDP}_{mean}},\:\forall\:i,
\end{equation}
where $\mbox{GDP}_{mean}\equiv\frac{\sum_i^N \mbox{GDP}_i}{N}$ is the average GDP for an observed year. The two quantities adjust the values of the countries GDPs for both the size of the network and the growth, and are a connected by a simple relation $\tilde{g}_{i}=N \cdot g_i$. We use the two quantities of the rescaled GDP throughout our analysis, mainly using $g_i$ for the reason that the quantity is bounded $0\le g_i\le1$ which coincides with our model.\\ 

In Fig. \ref{fig1} we plot the cumulative distribution of the rescaled GDP $\tilde{g}_{i}$ with $i$ indexing the countries for the different decades collected into our data set. What emerges is that the distributions of the rescaled GDPs can be described by log-normal distribution characterized by similar values of the parameters. The log-normal curve is fitted to all the values (from the different decades). This suggests that the rescaled GDPs are quantities which do not vary much with the evolution of the system, thus potentially representing the (constant) hidden macroeconomic fitness ruling the entire evolution of the system itself. This, in turn, implies understanding the functional dependence of the key topological quantities on the countries rescaled GDP.\\

\begin{figure*}[t!]
\caption{Comparison between observed (red points) degrees and strengths for the aggregated ITN in the 2000 snapshot. Right panel: degree $k_{i}$ versus normalized GDP $g_i$ and expected degree $\langle k_i\rangle$ versus normalized GDP $g_i$. Left panel: strength $s_{i}$ versus normalized GDP $g_i$ and expected strength $\langle s_i\rangle$ versus normalized GDP $g_i$.}
\centerline{\includegraphics[width=0.9\textwidth]{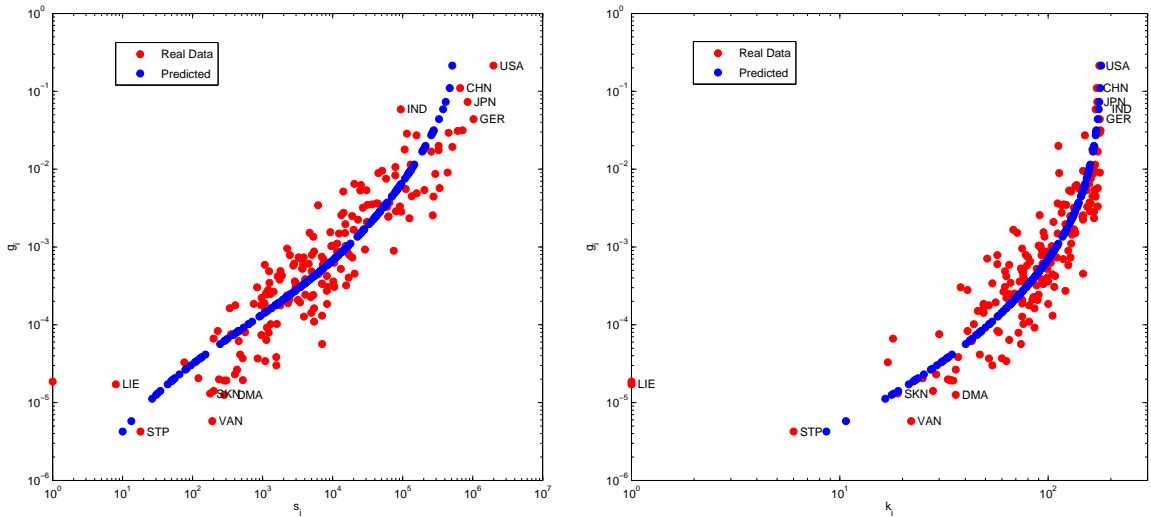}}
\label{fig2}
\end{figure*}

As already pointed out by a number of results \cite{Squartini3}, the topological quantities which play a major role in determining the ITN structure are the countries degrees (i.e. the number of their trading partners) and the countries strengths (i.e. the total volume of their trading activity). Thus, the first step to understand the role of the rescaled GDP in shaping the ITN structure is quantifying the dependence of degrees and strengths on it. Since we will now analyse each snapshot at a time (correction for size is not needed), here we will use the bounded rescaled GDP $g_i$. Moreover, this form of the rescaled GDP coincides with a bounded macroeconomic fitness value, which is consistent with the models presented in the next sections. 
To this aim, let us explicitly plot $k_i$ versus $g_i$ and $s_i$ versus $g_i$ for a particular decade, as shown in Fig. \ref{fig2}. The red points represent the relations between the two pairs of observed quantities for the 2000 snapshot. Interestingly, the rescaled GDP is directly proportional to the strength (in a log-log scale), thus indicating that \emph{the wealth of countries is strongly correlated to the total volume of trade they partecipate in}. Such an evidence provides the empirical basis for the definition of the gravity model, stating that \emph{the trade between any two countries is directly proportional to the (product of the) countries GDP}.

On the other hand, the functional dependence of the degrees on the $g_i$ values is less simple to decipher. Generally speaking, the relation is monotonically increasing and this means that countries with high GDP have also an high degree, i.e. are strongly connected with the others; coherently, countries characterized by a low value of the GDP have also a low degree, i.e. are less connected to the rest of the world. Moreover, while for low values of the GDP there seems to exist a linear relation (in a log-log scale) between $k_i$ and $g_i$, as the latter rises a saturation effect is observed (in correspondence of the value $k_{max}=N-1$), due to the finite size of the network under analysis. Roughly speaking, richest countries lie on the vertical trait of the plot, while poorest countries lie on the linear trait of the same plot: in other words, \emph{the degree of countries represents a purely topological indicator of the countries wealth}.

To sum up, Fig. \ref{fig2} shows that countries GDP plays a double role in shaping the ITN structure: first, it controls for the number of trading channels each country establishes; second, it controls for the volume of trade each country participates in, via the established connections.

The blue points in Fig. \ref{fig2}, instead, represent the relation between $\langle k_i\rangle$ versus $g_i$ and $\langle s_i\rangle$ versus $g_i$, where the quantities in brackets are the predicted values for degrees and strengths generated by our model, which we will discuss later.

\section{Null models}

In order to formalize the evidences highlighted in the previous section, a theoretical framework is needed. 
To this aim, we can make use of the \emph{exponential random graph} formalism (ERG in what follows).
Under this formalism, one ``generates'' a ensemble of random networks by maximizing the entropy of the ensemble.
However, the maximization is done under certain ``constraints'' which enforce certain properties of the random ensemble (expectations) to be equal specific observables that are measured in the real system.
Different maximum-entropy models enforce different constraints, different properties of the real network, and this corresponds to different probabilities and expectations of the models.\\ 

Here, we use the formulas defining the so-called \emph{enhanced configuration model} (ECM in what follows) which has been recently proposed as an improved model for the ITN reconstruction \cite{Squartini6}.
The ECM aims at reconstructing weighted networks, by enforcing the degree and the strength sequences simultaneously \cite{Squartini6}.
Degrees and strengths, respectively defined as $k_i(\mathbf{W})=\sum_{j\neq i}^N a_{ij}=\sum_{j\neq i}^N \Theta[w_{ij}],\:\forall\:i$ and $s_i(\mathbf{W})=\sum_{j\neq i}^N w_{ij},\:\forall\:i$, can be simultaneously constrained within into the ERG framework \cite{Squartini6}. From the perspective of network theory, specifying the countries degrees amounts to reproduce the binary structure of the ITN or, as previously said, its skeleton; on the other hand, specifying the countries strengths amounts to reconstruct the weight of each link. In economic terms, this amounts to retain two different kinds of information: the number of trading partners of each country \emph{and} the total volume of trade of each country.\\ 

Notice that previous attempts to infer the binary structure of the ITN from the information encoded into the strength sequence alone have led to the prediction of a largely homogeneous and very dense (sometimes fully connected) network, not compatible with the observed one. In other words, predicting the number of partners of a given country from the total volume of its trade leads to ``dilute'' the total trade of each country by distributing it to almost all other countries, dramatically overestimating the number of trading partners \cite{Squartini3}. 
This failure in correctly replicating the  purely topological projection of the real network is at the root of the bad agreement between expected and observed higher-order properties and makes it necessary to explicitly constrain the degree of each country. This evidence should lead us to reconsider the quantities traditionally used in economic models and the actual role played by them in explaining a given network structure. Particularly, one must add additional information regarding the topology of the network in order to reproduce the complex structure of the ITN. \\


As a result of constraining both degrees and strengths, the ECM predicts that a trade relation between countries $i$ and $j$ exists with a probability $p_{ij}$ equal to

\begin{equation}
\langle a_{ij} \rangle(\mathbf{x}, \mathbf{y})\equiv p_{ij}(\mathbf{x}, \mathbf{y})=\frac{x_ix_jy_iy_j}{1-y_iy_j+x_ix_jy_iy_j}
\label{eq_pp}
\end{equation}
and involves an expected volume of trade amounting to
\begin{equation}
\langle w_{ij} \rangle(\mathbf{x}, \mathbf{y})=\frac{p_{ij}(\mathbf{x}, \mathbf{y})}{1-y_iy_j}=\frac{x_ix_jy_iy_j}{(1-y_iy_j+x_ix_jy_iy_j)(1-y_iy_j)}.
\label{eq_ww}
\end{equation}

The unknown vectors $\mathbf{x}$ and $\mathbf{y}$ can be estimated according to the maximum-of-the-likelihood prescription \cite{Squartini5}, by solving the system of $2N$ coupled equations

\begin{equation}
k_i(\mathbf{W}^*)=\sum_{j\neq i}^N p_{ij}(\mathbf{x}^*, \mathbf{y}^*),\forall\:i\:\:\:\mbox{and}\:\:\:s_i(\mathbf{W}^*)=\sum_{j\neq i}^N \langle w_{ij}\rangle(\mathbf{x}^*, \mathbf{y}^*),\forall\:i\label{eq_ks}
\end{equation}
where $\mathbf{W}^*$ indicates the particular weighted network under analysis and $\mathbf{x}^*$ and $\mathbf{y}^*$ indicate the values of the Lagrange multipliers satisfying eqs.(\ref{eq_ks}). These parameters can be treated as fitness parameters, respectively controlling for the probability that a link exists and that its expected weight assumes a given value.

The application of the ECM to various real-world networks shows that the model can accurately reproduce the higher-order empirical properties of these networks \cite{Squartini5}. When applied to the ITN in particular, the ECM replicates both binary and weighted empirical properties, for different levels of disaggregation, and for several years \cite{Squartini6}.

\section{A GDP-driven model of the ITN}

Let us now make a step forward and check whether the hidden variables $x_i$ and $y_i$, which effectively reproduce the observed ITN \cite{Squartini6}, can be thought of as parameters having a clear (macro)economic interpretation. Let us start our analysis by first inspecting the relationship between the ECM statistics $k_i$ and $s_i$ and the hidden variables extracted from the model.

\begin{figure*}[t!]
\caption{
Comparison between observed relations of the degrees and strengths for the aggregated ITN in the 2000 snapshot. Right panel: degree $k_{i}$ versus normalized GDP $g_i$ (red points) and degree $k_i$ versus calculated fitness parameter $x_i$ (blue points). Left panel: strength $s_{i}$ versus normalized GDP $g_i$ (red points) and strength $ s_i$ versus calculated fitness parameter $y_i$ (blue points).}
\centerline{\includegraphics[width=0.9\textwidth]{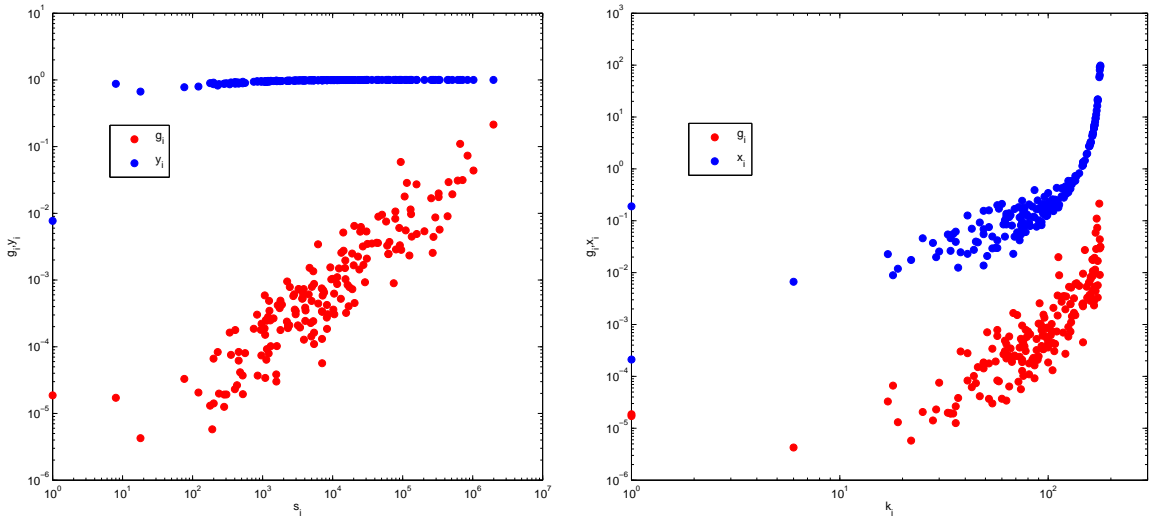}}
\label{fig3}
\end{figure*}

As Fig. \ref{fig3} shows, nodes degrees $k_i$ seems to be related to the quantities $x_i$ and $g_i$ through a very similar relationship; on the other hand, the functional relation between $s_i$ and $y_i$ appears to be less straightforward, showing a saturation effect in correspondence of the value $y=1$. In order to discover the mathematical form of these relations, let us repeat the analysis which led to Fig. \ref{fig3}, by plotting $x_i$ and $y_i$ versus $g_i$.

In Fig. \ref{fig4} we show the relationship between the two ECM parameters $x_i$ and $y_i$ and the rescaled GDP for each country of the ITN in the 2000 snapshot. Such quantities are strongly correlated, confirming the linear dependence between $x_i$ and $g_i$ and $y_i/(1-y_i)$ and $g_i$ respectively. The latter, in particular, is the simplest functional form guaranteeing the presence of the vertical asymptote emerging from the plot as $s_i$ versus $y_i$.

\subsection{The GDP as a macroeconomic fitness}

\begin{figure}[t!]
\caption{Comparison between the calculated $x_i$ and the rescaled GDP $g_i$ (left panel) and for the calculated $y_i/(1-y_i)$ and the relative GDP $g_i$ (right panel), for the aggregated ITN in the 2000 snapshot, together with a linear fit (black line).}
\centerline{\includegraphics[width=0.8\textwidth]{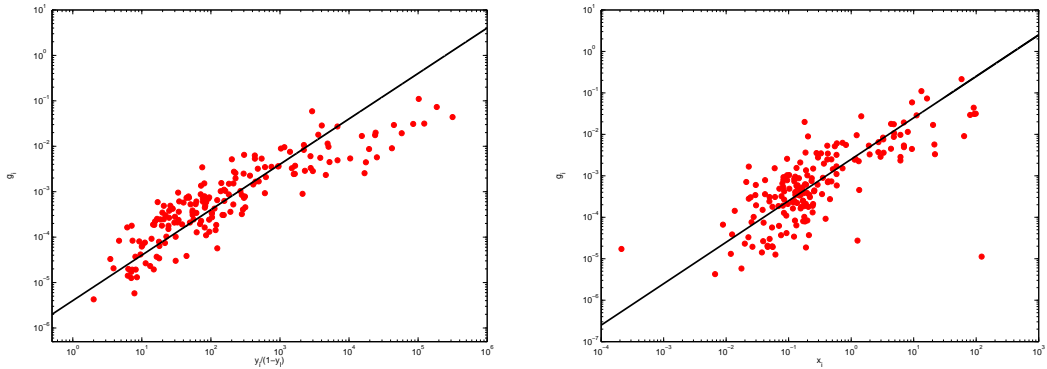}}
\label{fig4}
\end{figure}

Fig. \ref{fig4} seems to suggest that the fitness parameter $x_i$ satisfies a approximately linear relation with the relative GDP $g_i$, fitted by the curve

\begin{equation}
x_i=\sqrt{a}\cdot g_i
\label{eq_xx}
\end{equation}
where $\sqrt{a}$ is a parameter and $g_i=\frac{\mbox{GDP}_i}{\sum_i \mbox{GDP}_i}$.

By contrast, since the GDP is an unbounded quantity, while the fitness parameter $y_i$ is bounded between 0 and 1 (this is a mathematical property of the model \cite{Squartini5,Garlaschelli2}), the relation between $y_i$ and $g_i$ must be necessarily non-linear. A simple functional form for such a relationship is given by

\begin{equation}
y_i=\frac{b \cdot g_i^c}{1+b \cdot g_i^c}.
\label{eq_yy}
\end{equation}
Indeed, Fig. \ref{fig4} confirms that the above expression provides a very good fit to the data. 

These findings have two important consequences: first, they confirm that the GDP of world countries plays a double role, contributing to determine both the topological structure of the ITN and the amount of trade exchanges; second, since the relationships summed up by eqs.(\ref{eq_xx}) and (\ref{eq_yy}) hold true for each snapshot of the ITN in our data set, for each year we can insert eqs.(\ref{eq_xx}) and (\ref{eq_yy}) into eqs.(\ref{eq_pp}) and (\ref{eq_ww}) to obtain a GDP-driven model of the ITN structure for that year. While this was already expected on the basis of the results obtained by implementing simpler null models - constraining either the degree sequence alone (the \emph{binary configuration model}, or BCM \cite{Squartini3}) or the strength sequence alone (the \emph{weighted configuration model}, or WCM \cite{Squartini3}) - finding the appropriate way to explicitly combine these results into a unified description of the ITN has remained impossible so far.

\subsection{Reformulating the ECM as a ``two-step'' model}

It should be noted that eqs.(\ref{eq_xx}) and (\ref{eq_yy}) can be thought of as a particular case of a popular model among physicists, the so-called \emph{fitness model} \cite{fitness}, which prescribes to write the connection probability $p_{ij}$ between any two nodes $i$ and $j$ as a function of some intrinsic ``fitness'' characterizing each vertex. This observation leads to the identification of the fitness parameter with the GDP of countries, thus suggesting that, from a purely economic point of view, GDP is the only relevant quantity that must be taken into account in order to explain the observed structural patterns. Such a procedure, first adopted in \cite{Diego1} to study the purely binary structure of the ITN\footnote{In the BCM, the probability that any two nodes $i$ and $j$ are connected has the expression $p_{ij}^{BCM}=\frac{z_iz_j}{1+z_iz_j}$. The unknown parameters $\mathbf{z}$ can be numerically evaluated upon solving the system of $N$ equations $k_i(\mathbf{W}^*)=\sum_{j\neq i}^N p_{ij}^{BCM}(\mathbf{z}^*),\forall\:i$.} - and it has been shown that there is a very good agreement between the hidden variables $z_i$ which control solely for the degree sequence and the rescaled GDP - not only allows one to make predictions of the quantities of interest based on purely (country-specific) macroeconomic properties but also provides an algorithm to test the effectiveness of the chosen quantities in reproducing such observations. In fact, eqs.(\ref{eq_xx}) and (\ref{eq_yy}) could be, in principle, refined by further inserting any supposedly relevant macroeconomic quantity (as the geographic distances); however, their actual (macro)economic relevance would then be tested upon quantifying the actual fitting improvement.\\

At this point, it should be noted that we are arrived at two seemingly conflicting results. In fact we have explicitly stated that \emph{both} the BCM \emph{and} the ECM give a very good prediction for the binary topology of the ITN. However, the equations 
which specify the connection probability $p_{ij}$ in the two models are significantly different. Thus, these findings make us expect that the values of the probabilities of connection for each single pair of nodes become comparable in these two models, despite the different mathematical expressions. \cite{Almog} shows the comparison of the two probability matrices $\{p_{ij}^{BCM}\}$ and $\{p_{ij}^{ECM}\}$ are in fact very similar.
This in turn, enables us to greatly simplify the equations defining the ECM, by replacing the expression for the $p_{ij}$ coefficients provided by the ECM with that provided by the BCM. If we denote the new probability coefficients with $p_{ij}^{ts}$, ``$ts$'' standing for ``two-step'' (the reason will be clear in a moment), eqs.(\ref{eq_pp}) and (\ref{eq_ww}) can be naturally rewritten as 

\begin{eqnarray}
\langle a_{ij}\rangle^{ts}(\mathbf{z})&\equiv& p_{ij}^{ts}(\mathbf{z})=\frac{z_iz_j}{1+z_iz_j},\label{eq_pts}\\
\langle w_{ij}\rangle^{ts}(\mathbf{z}, \mathbf{y})&=&\frac{p_{ij}^{ts}(\mathbf{z})}{1-y_iy_j}.\label{eq_wts}
\end{eqnarray}
where, now, the unknown vector $\mathbf{z}$, and therefore the $p^{ts}_{ij}$ coefficients, can be determined by solving a system of equations formally analogue to the one defining the BCM, i.e. $k_i(\mathbf{W}^*)=\sum_{j\neq i}^N p_{ij}^{ts}(\mathbf{z}^*),\forall\:i$. In this simplified model the connection probabilities no longer depend on the strengths as in the original ECM, while the weights still do. In other words, we have decoupled the structural part of the system of equations defining the ECM from the remaining one, providing a simpler set of equations to solve. This, in turn, implies that we can specify the model via a ``two-step'' procedure according to which 1) we first solve the $N$ equations determining the $p_{ij}^{ts}$, upon constraining the nodes degrees only and 2) then evaluate the remaining parameters determining $\langle w_{ij}\rangle^{ts}$ through the ECM. For this reason, we denote the model as the ``two-step'' model (TS hereafter).

The TS model inherits the functional form of the link-specific distribution of weights from the ECM:

\begin{equation}
q_{ij}^{ts}(w_{ij})=\frac{(z_iz_j)^{a_{ij}} (y_iy_j)^{w_{ij}-a_{ij}}(1-y_iy_j)^{a_{ij}}}{1+z_iz_j}.
\label{prob}
\end{equation}

It is instructive to rewrite eq.(\ref{prob}) as a product of two different factors, i.e. as $q_{ij}^{ts}(w_{ij})=\left[\frac{(z_iz_j)^{a_{ij}}}{1+z_iz_j}\right]\cdot(y_iy_j)^{w_{ij}-a_{ij}}(1-y_iy_j)^{a_{ij}}$ to highlight the random processes behind the formation of each link. As a first step, one implements a Bernoulli trial with probability $p_{ij}^{ts}$ in order to determine whether a link connecting $i$ and $j$ is created or not. The second part of our algorithm can be interpreted as a drawing from a geometric distribution, with parameter $y_iy_j$: if a link (or, equivalently, a unitary weight) is indeed established, a second random process determines whether the weight of the same link is increased by another unit (with probability $y_iy_j$) or whether the process stops (with probability $1-y_iy_j$). Iterating this procedure to determine the probability of obtaining weights of higher values leads precisely to eq.(\ref{prob}). As a consistency check, one can explicitly calculate the expected weight $\langle w_{ij}\rangle^{ts}$ for the nodes pair $i$-$j$ through the formula $\sum_{w=0}^{+\infty}w\cdot q_{ij}^{ts}(w)$, which correctly leads to eq.(\ref{eq_wts}).\\

In more economic terms, the analysis of the ITN clearly proves that a substantial difference exists between establishing a new trade relation and reinforcing an existing one by rising the exchanged amount of goods of e.g. ``one unit'' of trade. These two processes are described, respectively, by the coefficients $p_{ij}^{ts}$ and $y_iy_j$. In order to understand which one is more probable, we can study the behavior of the ratio $p_{ij}^{ts}/(y_iy_j)$ for each pair of countries. In fact, whenever $p_{ij}^{ts}/(y_iy_j)>1$ countries $i$ and $j$ would probably establish a new trade relation quite easily, however experiencing a certain resistance to reinforce it. On the other hand, whenever $p_{ij}^{ts}/(y_iy_j)<1$ countries $i$ and $j$ would experience a certain resistance to start trading; however, in the case such a relation were established, it would represent a channel with relatively low ``friction'', inducing the involved parteners to strengthen it. 

Before analysing the case $p_{ij}^{ts}/(y_iy_j)=1$ let us rewrite it as $\frac{z_iz_j}{y_iy_j}(1-y_iy_j)=1$. The expression at the first member appears also in eq.(\ref{prob}) which, in fact, can be restated in the following way: $q_{ij}^{ts}(w_{ij})=\left[\frac{z_iz_j}{y_iy_j}(1-y_iy_j)\right]^{a_{ij}}\frac{(y_iy_j)^{w_{ij}}}{1+z_iz_j}$. Imposing the first factor to be equal to 1 implies reducing eq.(\ref{prob}) to $q_{ij}^{ts}(w_{ij})=(y_iy_j)^{w_{ij}}(1-y_iy_j)$, i.e. to the WCM probability distribution. This model does not discriminate between the first link and the subsequent ones, reducing \emph{tout court} $q_{ij}^{ts}(w_{ij})$ to a simple geometric distribution: thus, the failure of the WCM in reproducing the observed properties of the ITN lies precisely in its incapability to give the right importance to the very first link, treating it as a simple unit of weight and not as the channel making the trade exchanges possible.

\subsection{A GDP-driven model of the ITN}

Eqs.(\ref{eq_pts}) and (\ref{eq_wts}) provide the expressions into which we can input the vector of fitness parameters $g_i,\:\forall\:i$, according to the prescriptions of eqs.(\ref{eq_xx}) and (\ref{eq_yy}). As a result, we obtain the following formulas that mathematically characterize our GDP-driven specification of the TS model: 

\begin{eqnarray}
\langle a_{ij}\rangle^{ts}(a)&\equiv&p_{ij}^{ts}(a)=\frac{a\cdot g_ig_j}{1+a\cdot g_ig_j},\\
\langle w_{ij}\rangle^{ts}(a, b, c)&=&p_{ij}^{ts} \frac{(1+b\cdot g_i^c)(1+b\cdot g_j^c)}{(1+b\cdot g_i^c+b\cdot g_j^c)}.
\label{Twostep1}
\end{eqnarray}

Eqs.(\ref{Twostep1}) can be used to reverse the approach used so far: rather than determining the $2N$ free parameters either of the ECM ($\mathbf{x}$ and $\mathbf{y}$) or of the TS model ($\mathbf{z}$ and $\mathbf{y}$), upon constraining degrees and strengths to their observed values, we can now use the knowledge of the GDP of all countries to obtain a model that only depends on the three parameters $a$, $b$, $c$. Since the model consists of two subsequent steps, we can first assign a value to the parameter $a$ and, only once $a$ is set, fit the remaining parameters $b$ and $c$.

Parameter $a$ can be determined quite easily. In fact, following \cite{Diego1,Diego2}, the value of $a$ can be chosen as the one ensuring that the density of connections is reproduced, i.e.

\begin{equation}
L=\sum_i^N\sum_{j\neq i}^N\frac{a\cdot g_ig_j}{1+a\cdot g_ig_j};
\label{fcm}
\end{equation}
such a prescription overcomes the limitation of econometric models (as the gravity model) in failing to predict the right density of connections, allowing us to fix it from the very beginning. Notice that satisfying eq.(\ref{fcm}) is equivalent to maximizing the likelihood function of the fitness model, as shown in \cite{Garlaschelli}.

\begin{figure*}[t!]
\caption{Comparison between the observed properties (red points), the corresponding ensemble averages of the GDP-driven ``two-step'' model (blue points)  of the aggregated ITN in the 2000 snapshot. Left panel: average nearest neighbors degree $k^{nn}_i$ versus degree $k_i$. Right panel: average nearest neighbors strength $s^{nn}_i$ versus strength $s_i$.}
\centerline{\includegraphics[width=0.9\textwidth]{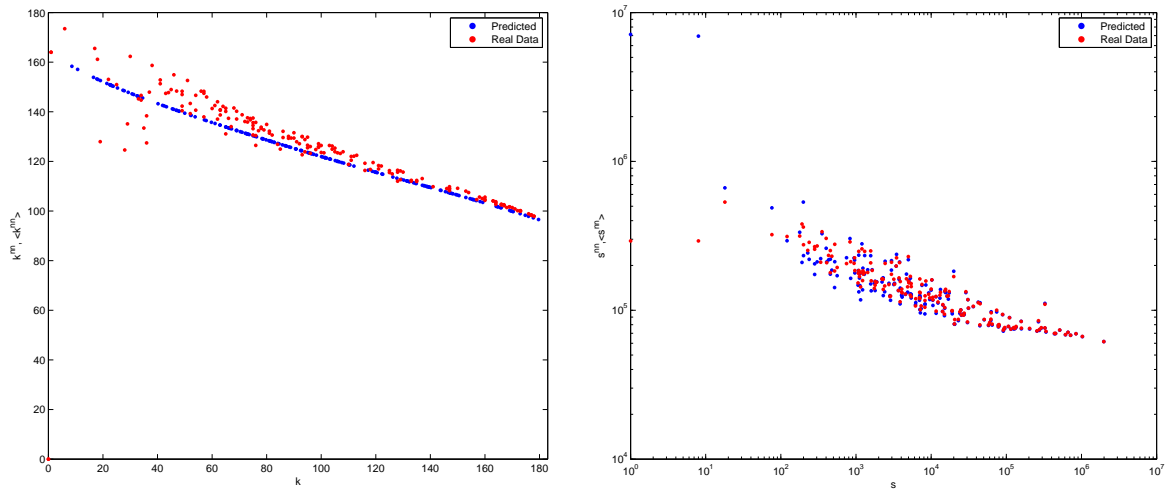}}
\label{fig5}
\end{figure*}

Fixing the values of $b$ and $c$ is slightly more complicated. In fact, we could imagine to impose a similar condition, as constraining the total weight $W$ of the network. However, since the TS model uses approximate expressions, rather than those of the ECM, maximizing the likelihood function in the second step of the model no longer coincides with the desired condition $\langle W\rangle=W$. Similarly, extracting the parameters from the fit shown in Fig. \ref{fig3} does not preserve the total weight of the network. However, in absence of any \emph{a priori} preference, we chose the latter procedure, due to its relative numerical simplicity with respect to the former one.\\

In Fig. \ref{fig4} we show a comparison between the higher-order observed properties of the ITN in 2000 and their expected counterparts predicted by the GDP-driven TS model (the mathematical expressions of these properties are provided in Appendix). As a baseline comparison, we also show the predictions of the GDP-driven WCM model with continuous weights proposed in \cite{Fronczak}, which coincides with a simplified version of the gravity model. 

Naturally, as expected, the predictions in Fig. \ref{fig4} are more noisy than the ECM predicted values (the TS model makes use of three parameters only, while the ECM is defined by $2N$ parameters): this is due to the fact that eqs.(\ref{eq_xx}) (and the corresponding BCM equation) and (\ref{eq_yy}) describe fitting curves rather than exact relationships. However, as a general comment, the GDP-driven TS model reproduces the empirical trends very well; most importantly, our model performs significantly better than the GDP-driven WCM in replicating both binary and weighted properties. Again, the drawback of these models lies in the fact that they predict a fully connected topology and a relatively homogeneous network. More specifically, the plot of the average nearest neighbor strengths, $s^{nn}$, predicted by our model is slightly shifted with respect to the observed points. This effect is due to the fact that, as we mentioned, the total weight of the network $W$ (hence the average trend of $s^{nn}$) is only approximately reproduced by our model, as a consequence of the simplification leading from the ECM to the TS model. Our findings are also robust over the entire time span of our data set. We can therefore conclude that the ECM model, as well as its simplified TS variant, can be successfully turned into a fully GDP-driven model that simultaneously reproduces both the topology and the weighted structure of the ITN.

\section{Conclusion}

In this paper we have demonstrated the capabilities of a novel GDP-driven model which successfully reproduces both the binary and weighted properties of the ITN. The model uses the GDP of world countries as a sort of macroeconomic fitness that in turns determine the probabilities for the formation of the network links. The use of the GDP as a macroeconomic fitness parameter is motivated in the first section, where we show the extent to which this quantity is entangled with the first order, country-specific, properties of the network. The model also represent an improvement in the reconstruction ability of a network, by extending it to both the binary ant the weighted representations.\\

The success of the TS model has an important interpretation. We recall that the effect of the approximation leading from the ECM to the TS model lies in the fact that the connection probability $p_{ij}^{ts}$ can be estimated separately from the weights $\langle w_{ij}\rangle^{ts}$, using either the knowledge of the degree sequence - if eq.(\ref{eq_pts}) is used - or that of the GDPs and total number of links - if eq.(\ref{Twostep1}) is used. By contrast, the estimation of the expected weights cannot be carried out separately, as it requires the evaluation of the connection probability $p_{ij}^{ts}$.
This asymmetry implies that the topology of the ITN can be successfully inferred without any information about the weighted properties, while the weighted structure cannot be inferred without any topological information.
This effect is thus the origin of the limitation of ``purely weighted' models, such as the Gravity Model, which focus on trade volumes while disregarding the connectivity of countries. The TS model provides a mathematical explanation for this otherwise puzzling effect observed in the ITN.


\begin{table*}[t!]
\centering
\begin{tabular}{|c|c|c|}
\hline
\mbox{\textbf{Empirical properties}} & \mbox{\textbf{Expected properties under the ECM}} & \mbox{\textbf{Expected properties under the TS}}\\
\hline
\hline
$a_{ij}$ & $\langle a_{ij}\rangle=p_{ij}=\frac{x_{i}x_{j}y_{i}y_{j}}{1-y_{i}y_{j}+x_i x_jy_{i}y_{j}}$ & $\langle a_{ij}\rangle=p_{ij}^{ts}=\frac{z_{i}z_{j}}{1+z_{i}z_{j}}$\\
\hline
$k_{i}=\sum_{j\ne i}a_{ij}$ & $\langle k_{i}\rangle=\sum_{j\ne i}p_{ij}=k_i$ & $\langle k_{i}\rangle^{ts}=\sum_{j\ne i}p_{ij}^{ts}$\\
\hline
$k_{i}^{nn}=\frac{\sum_{j\ne i}a_{ij}k_{j}}{k_{i}}$ & $\langle k_{i}^{nn}\rangle=\frac{\sum_{j\ne i}p_{ij}k_{j}}{k_{i}}$ & $\langle k_{i}^{nn}\rangle^{ts}=\frac{\sum_{j\ne i}p_{ij}^{ts}\langle k_{j}\rangle^{ts}}{\langle k_{i}\rangle^{ts}}$\\
\hline
$c_{i}=\frac{\sum_{j\ne i}\sum_{k\ne i,j}a_{ij}a_{jk}a_{ki}}{
k_i(k_i-1)}$ & $\langle c_{i}\rangle=\frac{\sum_{j\ne i}\sum_{k\ne i,j}p_{ij}p_{jk}p_{ki}}{\sum_{j\ne i}\sum_{k\ne i,j}p_{ij}p_{ik}}$ & $\langle c_{i}\rangle^{ts}=\frac{\sum_{j\ne i}\sum_{k\ne i,j}p_{ij}^{ts}p_{jk}^{ts}p_{ki}^{ts}}{\sum_{j\ne i}\sum_{k\ne i,j}p_{ij}^{ts}p_{ik}^{ts}}$\\
\hline
\hline
$w_{ij}$ & $\langle w_{ij}\rangle=\frac{p_{ij}}{1-y_{i}y_{j}}$ & $\langle w_{ij}\rangle^{ts}=\frac{p_{ij}^{ts}}{1-y_{i}y_{j}}$\\
\hline
$s_{i}=\sum_{j\ne i}w_{ij}$ & $\langle s_{i}\rangle=\sum_{j\ne i}\langle w_{ij}\rangle$ & $\langle s_{i}\rangle^{ts}=\sum_{j\ne i}\langle w_{ij}\rangle^{ts}$\\
\hline
$s_{i}^{nn}=\frac{\sum_{j\ne i}a_{ij}s_{j}}{k_{i}}$ & $\langle s_{i}^{nn}\rangle=\frac{\sum_{j\ne i}p_{ij}s_{j}}{k_{i}}$ & $\langle s_{i}^{nn}\rangle^{ts}=\frac{\sum_{j\ne i}p_{ij}^{ts}\langle s_{j}\rangle^{ts}}{\langle k_{i}\rangle^{ts}}$\\
\hline
\end{tabular}
\caption{Mathematical expressions for the empirical and expected properties of the undirected representation of the ITN.\label{tab_b}}
\end{table*}

\appendix
\subsection*{Appendix: higher-order properties of the undirected representation of the ITN}

Table \ref{tab_b} gives a summarized description of the binary and weighted network quantities analysed in this paper. Specifically, it both shows their analytical definition and the corresponding expected value under the ECM and the GDP-driven TS model.

Let us recall that a weighted undirected network can be represented through a square matrix $\mathbf{W}$, where the specific entry $w_{ij}$ represents the edge weight between country $i$ and country $j$. The binary representation of the network, encoded into the matrix $\mathbf{A}$, is straightforwardly obtained upon defining $a_{ij}\equiv\Theta[w_{ij}]$.

The \emph{degree} and the \emph{strength} of a given node, respectively defined as $k_i(\mathbf{W})=\sum_{j\neq i}^N a_{ij}=\sum_{j\neq i}^N \Theta[w_{ij}],\:\forall\:i$ and $s_i(\mathbf{W})=\sum_{j\neq i}^N w_{ij},\:\forall\:i$, are first-order properties, describing the neighborhood of the node itself and, specifically, the number of its first neighbors (i.e. the other nodes sharing a direct connection with it) and its total volume.

Exploring the topological properties of more distant nodes (i.e. the neighbors of the neighbors) implies considering longer pathways starting from node $i$. The simpler second-order properties that can be defined are the \emph{average nearest neighbors degree}, $k_i^{nn}$, i.e. the arithmetic mean of the degrees of the neighbors of node $i$ and the \emph{average nearest neighbors strength}, $s_i^{nn}$, i.e. the arithmetic mean of the strengths of the neighbors of node $i$. Once plotted versus the corresponding node degree (strength), the $k_i^{nn}$ ($s_i^{nn}$) provides information on the tendency of nodes degrees (strengths) to be either positively or negatively correlated. In economic terms, the $k^{nn}$ quantifies the tendency of strongly connected countries to trade with strongly connected partners as well. 

Another important feature of complex networks concerns the tendency of nodes to cluster together. It can be quantified through the clustering coefficient, $c_i$, which measures the percentage of closed triangles node $i$ is part of. In economic terms, the clustering coefficient quantifies the tendency of countries to form small communities and, at a more general level, the hierarchical character of the ITN structure.\\

The measured properties of the real network need to be compared with the different models predictions. The expected values can be obtained by simply replacing $a_{ij}$ with the probability coefficients $\langle a_{ij}\rangle$ predicted by the different models (e.g. $\langle a_{ij}\rangle=\frac{z_iz_j}{1+z_iz_j}=p_{ij}^{ts}$ for the TS, $\langle a_{ij}\rangle=\frac{x_ix_jy_iy_j}{1-y_iy_j+x_ix_jy_iy_j}$ for the ECM, etc.) and $w_{ij}$ with $\langle w_{ij}\rangle$ (e.g. $\langle w_{ij}\rangle=\frac{p_{ij}^{ts}}{1-y_iy_j}$ for the TS, etc.). Whenever considering the GDP-driven TS model, the mathematical expressions for $\langle a_{ij}\rangle$ and $\langle w_{ij}\rangle$ are the ones illustrated by eqs.(9) and (\ref{Twostep1}).

\end{document}